
\input harvmac

\noblackbox
\Title{\vbox{\baselineskip12pt{\hbox{CTP-TAMU-36/92}}%
{\hbox{hepth@xxx/9205083}}}}
{\vbox{\centerline{Solitons Reduced From Heterotic Fivebranes}}}

\centerline{HoSeong ~La\footnote{$^*$}{%
e-mail address: hsla@phys.tamu.edu, hsla@tamphys.bitnet}   }

\bigskip\centerline{Center for Theoretical Physics}
\centerline{Texas A\&M University}
\centerline{College Station, TX 77843-4242, USA}
\vskip 1in
In view of the expectation that the solitonic sector of the lower dimensional
world may be originated from the solitonic sector of string theory,
various solitonic solutions are reduced from the
heterotic fivebrane solutions in the ten-dimensional heterotic
string theory. These solitons in principle can appear after
proper compactifications,
{\it e.g.} toroidal compactifications.

\Date{5/92} 

\def\la{\lambda}
\def\half{{\textstyle{1\over 2}}}
\def\d{{\rm d}}
\def\e{{\rm e}}
\def\pa{\partial}

\def\eps{\epsilon}

\font\cmss=cmss10 \font\cmsss=cmss10 scaled 833
\def\IZ{\relax\ifmmode\mathchoice
{\hbox{\cmss Z\kern-.4em Z}}{\hbox{\cmss Z\kern-.4em Z}}
{\lower.9pt\hbox{\cmsss Z\kern-.4em Z}}
{\lower1.2pt\hbox{\cmsss Z\kern-.4em Z}}\else{\cmss Z\kern-.4em Z}\fi}
\def\tr{{\rm tr}}
\def\Tr{{\rm Tr}}
\def\cos{{\rm cos}}
\def\sin{{\rm sin}}
\def\CS{{\cal S}}
\def\CO{{\cal O}}
\def\IR{\relax{\rm I\kern-.18em R}}
\font\cmss=cmss10 \font\cmsss=cmss10 scaled 833

\vfill\eject
\newsec{Introduction}

The study of solitons has long been pursued in various aspects by physicists as
well as by mathematicians. In general, it involves the investigation of
nonlinear evolution equations. In one spatial dimensional case, one can mainly
deal with ordinary differential equations, which makes the situation relatively
easier.
 But in higher dimensional cases, since
one has to solve partial differential equations, the whole set of solutions are
almost impossible to find. Nevertheless many interesting solutions are known
and now string theory is not an exception anymore.

Lately the structures of the classical solitonic solutions of string theory
have been actively investigated\ref\Fbrev{For recent reviews, see
M.J.~Duff and J.X. Lu, ``A Duality between Strings and Fivebranes,''
Class. Quant. Grav. {\bf 9}  (1992) 1;
C.G. Callan, J.A.~Harvey and A.~Strominger ``Supersymmetric String Solitons ,''
Chicago preprint, EFI-91-66 (1991); and references therein.}.
In particular, the heterotic fivebrane solution conjectured by Duff\ref\Duf{%
M.J. Duff, Class. Quan. Grav. {\bf 5} (1988) 189\semi
M.J. Duff, in {\it Superworld II}, ed. by A. Zichichi (Plenum, New York, 1990).
}\ and constructed by
Strominger\ref\Stro{A. Strominger, Nucl. Phys. {\bf B343} (1990) 167.}\
is exceptionally interesting in the sense that it is dual
to the fundamental string in a generalized sense of the electric-magnetic
duality\foot{This duality which interchanges
Noether charge (e.g. electric charge)
and topological charge (e.g. monopole charge) is in principle the foundation
for
the Montonen-Olive conjecture\ref\MoOl{C. Montonen and D. Olive, Phys. Lett.
{\bf 72B} (1977) 117.}, which is yet to be confirmed rigorously.}.
However, most of the solutions known so far are rather ten-dimensional
solutions
so that their fate in 4-d space-time after compactification is still elusive.

Thus it is important to address a
question that what would be the implications of the
physics of the fivebrane in ten-dimension on the
physics in four-dimensional space-time
after some proper
 compactification\foot{Also with broken space-time supersymmetry ultimately,
but such solutions are not known yet.}.
Some speculations were given by Strominger, too\Stro.

There may be some physical consequences
due to the above duality. For example, the origin of
the electric-magnetic duality
in four-dimension might be such a string-fivebrane duality in ten-dimension.
In other words,
the monopole solution in four-dimension might be related to the fivebrane
solution in ten-dimension. This aspect was already advocated by Harvey and
Liu\ref\HaLi{J.A. Harvey and J. Liu, Phys. Lett. {\bf 268B} (1991) 40.}.
The dynamical similarities between these two systems are investigated
classically in ref.\ref\Laorb{R. Khuri and H.S. La, Texas A\&M preprint,
CTP-TAMU-95/91, -98/91 (1991).}.
Furthermore, we could conjecture that the solitonic sector in four-dimension is
originated from the solitonic sector of ten-dimension.

In this paper, as a first step toward such structures  in the
solitonic sector of string theory, we attempt to investigate how these
solutions
appear in the $(1+3)$-dimensional subspace of
the $(1+9)$-dimensional space-time. Perhaps this could provide
some clues to consistent compactifications of the fivebrane system.
By proper coordinate redefinitions and the dimensional reduction imposing the
Killing symmetries,
we make the fields in this subspace
independent from the rest of the space so that they are more or less
dimensionally reduced solutions. In particular most of new solutions we present
here are not based on the instanton background in the transverse
space\ref\Lasg{H.S. La, Texas A\&M preprint, CTP-TAMU-96/91
(hepth@xxx/9205035).}.

Later, some remarks on the compactified solutions will also be given, whose
detail will be presented elsewhere. Also we can attempt to analyze the motion
of strings in the fivebrane geometry inside $(1+3)$-dimensional space-time,
using the metric suggested here\ref\Laf{H.S. La, in preparation.}.

This paper is organized as follows: In sect. 2, we review the derivations of
the basic fivebrane solutions. Then in sect.3, the hyperbolic monopoles and the
neutral fivebrane solutions are derived by imposing rotational symmetries.
In sect.4 the (Euclidean) sine-Gordon solitons are used as sources of new
fivebrane solutions and in sect.5 the Euclidean anolgues of the $\Phi^4$ kinks
are used to derive new fivebranes. In sect.6 using the conformal mapping we
attempted to analyze the instanton case and finally in sect.7 some speculation
on compactified fivebrane solutions is presented.

\newsec{Heterotic Fivebranes}

A fivebrane is a five dimensional extended object and the existence of such a
higher dimensional object is in some sense surprising. Nevertheless, such a
solution exists in string theory.
First of all,
let us review the derivations given in refs.\Stro\ref\CaHaSt{C.G.
Callan, J. Harvey and A.
Strominger, Nucl. Phys. {\bf B359} (1991) 611.}.

The heterotic fivebrane is a solution to the equations of the supersymmetric
vacuum for the heterotic string
\eqn\suei{\delta\psi_M=\left(\pa_M+{\textstyle {1\over 4}}
\Omega_{MAB}\Gamma^{AB}\right)\eps=0,}
\eqn\sueii{\delta\lambda=\left(\Gamma^A\pa_A\phi+{\textstyle{1\over 6}}H_{AMC}
\Gamma^{ABC}\right)\eps=0,}
\eqn\sueiii{\delta\chi=F_{AB}\Gamma^{AB}\eps=0,}
where $\psi_M,\ \lambda$ and $\chi$ are the gravitino, dilatino and gaugino,
and the generalized connection is given by
\eqn\gencon{\Omega_{M}^{AB}\equiv\omega_M^{AB}-H_M^{AB},}
where $\omega$ is the usual spin connection.
Now $H$ satisfies the following anomaly equation:
\eqn\anoe{\d H=\alpha' \left(\tr R\wedge R-{\textstyle{1\over 30}}\Tr
	F\wedge F\right)+\CO(\alpha'^2).}
In the above we have properly rescaled all the field variables and that the
string coupling $g_s=\e^{-\phi}$ and $\alpha'$ are only independent
couplings. In the heterotic string theory $\alpha'$ is proportional to
$\kappa^2/g_{{\rm YM}}$, where $\kappa$ is the gravitational coupling
constant.

The corresponding low-energy effective action for the heterotic string is
\eqn\eef{\CS={1\over \kappa^2}\int\d^{10}x{\sqrt{-g}}\e^{2\phi}
\left(R+4\pa_\mu\phi\pa^\mu\phi-{\textstyle{1\over 3}}H^2-
{\textstyle{1\over 30}}\alpha'\Tr F^2 +\cdots\right),}
where the dots include the fermionic part of the action that are not relevant
for our purpose now.

In $(1+9)$-dimension we have Majorana-Weyl fermions, which decompose down to
chiral spinors according to SO(1,9)$\supset$SO(1,5)$\otimes$SO(4) for
$M^{1,9}\to M^{1,5}\times M^4$ decomposition.
For such spinors the dilatino
equation eq.\sueii\ is satisfied by
\eqn\ei{H_{\mu\nu\la}=\pm\eps_{\mu\nu\la\sigma}\pa^\sigma\phi,}
where $\mu,\nu,...$ are indices for the transverse space
$M^4$ and $\phi=\phi(x^\mu)$, while we shall use indices $a, b,...$ for
$M^{1,5}$. The dilaton itself is determined by solving the anomaly equation.

Then other equations are solved by constant chiral spinors $\eps_{\pm}$ and the
metric
\eqn\eii{g_{ab}=\eta_{ab},\ \ \ g_{\mu\nu}=\e^{-2\phi}\delta_{\mu\nu}}
such that
\eqn\eiii{\eqalign{\delta\psi_\mu&=\left(\nabla_\mu+{\textstyle{1\over 2}}
\Gamma_{\mu\nu}\pa^\nu\phi\right)\eps_\pm=\pa_\mu\eps_\pm=0,\cr
\delta\psi_a&=\nabla_a\eps_\pm=\pa_a\eps_\pm=0,\cr}}
and
\eqn\eiv{\delta\chi=F^\pm_{\mu\nu}\Gamma^{\mu\nu}\eps_\pm=
-F^\pm_{\mu\nu}\Gamma^{\mu\nu}\eps_\pm=0,}
where eq.\eiv\ is achieved using the  instanton configuration for the
(anti)self-dual YM equation in the flat Euclidean space $\IR^4$
\eqn\eins{F^\pm_{\mu\nu}=\pm\half\eps_{\mu\nu\rho\sigma}
F_{\rho\sigma}^\pm}
for an
SU(2) subgroup of $E_8\times E_8$ or SO(32).

Solutions of eq.\eins\ are basic ingredients to build fivebrane solutions.
For example, the instanton solutions lead to the Strominger's fivebrane
solutions. There are two relevant fivebrane solutions. One is the
``gauge'' solution and the other is the ``symmetric'' solution.
We shall first derive the former, then the latter.
In this case $\phi=\phi(r^2)$ now,
i.e. no angular dependence, where $r^2=\sum(x^\mu)^2$.
With a finite instanton scale size $\lambda$, from eqs.\anoe\ei\ we obtain
\eqn\eexpf{\e^{-2\phi}=\e^{-2\phi_0}+8\alpha'{(r^2+2\lambda^2)\over
(r^2+\lambda^2)^2},}
where $\phi_0$ is the value of the dilaton at spatial infinity.
Thus we have a fivebrane living in $M^{1,5}$ which is a point-like
object in $M^4$. This is the gauge solution.

This gauge solution is valid only for $\la\gg\sqrt{\alpha'}$. Nevertheless,
there is another fivebrane solution with $\la=0$, which is called
the elementary fivebrane or the ``neutral" solution\ref\DuLu{M. Duff
and J.X. Lu,  Nucl. Phys. {\bf 354} (1991) 141.}\CaHaSt.
For this neutral solution the YM fields vanishes
and as a result,
the dilaton is given by the same form as following ``symmetric" solution.

The symmetric solution can be
derived by setting the RHS of eq.\anoe\ zero.
 Then compared to the gauge
solution we derived before, the differences are
\eqn\esxpf{\e^{-2\phi}=\e^{-2\phi_0}+{Q\over r^2},\ \ Q=n\alpha',}
and
\eqn\eyms{F_{\mu\nu}=R_{\mu\nu}(\Omega),}
where $F$ and $R$ are both self-dual. This symmetric solution is known to be
exact\ref\CaHaStII{C.G. Callan, J. Harvey and A.
Strominger, Nucl. Phys. {\bf B367} (1991) 60.}.

Note that this solutions satisfy the  scale symmetry
\eqn\escs{\eqalign{\phi &\to\phi+\ln\sigma,\cr
                r &\to \sigma^{-1}r,\cr
                \la &\to \sigma^{-1}\la,\cr}}
where $\sigma$ is a constant.
The $\la=0$ case is not related to the $\la\neq 0$ case in terms of this scale
symmetry, but it retains a similar scale symmetry without the last property.

Now we would like to call the reader's attention to the fact
that any solution of eq.\eins\ in principle leads to a fivebrane
solution, as long as the anomaly equation eq.\anoe\ provides a nontrivial
solution for the dilaton. In
particular many lower dimensional solutions to the self-dual YM
equation are known\ref\rWa{R.S. Ward,
``Integrable Systems in Twistor Theory,"
in {\it Twistors in Mathematics and Physics}, (Cambridge Univ. Press,
1990).}\ so that in principle we can relate all these solitonic solutions to
the heterotic fivebranes.

\newsec{Rotationally Symmetric Cases}

The Bogomol'nyi equation can be reduced from the SDYM equation by requiring
one Killing symmetry along one of the cartesian coordinates. Then the BPS
monopole solution is the SO(3) rotational symmetric solution, which is used in
ref.\HaLi. If we claim other rotational symmetries in some subspaces of the
transverse space, we can obtain other type of fivebrane solutions closely
related to the solutions derived in the previous section.

First, there is a $S^1$-invariant instanton solution.
The basic observation is that for $(x^\mu)=(x^1, x^2, u, v)$ we can
introduce a set of cylindrical coordinates
\eqn\ecor{u=\rho\cos\theta,\ \ \ v=\rho\sin\theta,}
then the metric for $M^4$ can be rewritten as
\eqn\emtr{ds_4^2=\e^{-2\phi}\left( (dx^1)^2+(dx^2)^2+(d\rho)^2
+\rho^2(d\theta)^2\right),}
Since $r^2=(x^1)^2+(x^2)^2+\rho^2$, now the dilaton $\phi$ will be defined in
terms of the coordinates $x^1, x^2, \rho$.

In this coordinate system the YM vector fields can be identified as
\eqn\ymv{A_\mu dx^\mu=A_1 dx^1+A_2 dx^2+ A_\rho d\rho + \rho A_\theta d\theta,}
Now we can adopt the Killing reduction of the self-dual YM
equation to the lower dimensional integrable systems\rWa.
If we require a Killing symmetry such that the YM vector fields do not depend
on the $\theta$-coordinate, i.e $\pa_\theta A_\mu=0$,
and define a scalar field as $\Phi\equiv A_\theta$, then the self-dual YM
equation becomes
\eqn\reym{F_{12}=D_\rho\Phi,\ \ F_{\rho 1}=D_2\Phi,\ \ F_{2\rho}=D_x\Phi.}
This is a set of Bogomol'nyi equations except the condition $\rho\geq 0$.
Note that the above procedure is in fact equivalent to getting the hyperbolic
monopoles from the self-dual YM equation using the conformal equivalence of
$R^4-R^2\sim S^1\times H^3$\ref\atiyah{M.F.  Atiyah, Comm. Math. Phys. {\bf 93}
(1984) 437.}, where $H^3$ is the hyperbolic space of an upper half plane
$(x^1, x^2, \rho)$.
Thus we basically recover a fivebrane with an extra $S^1$-rotational invariance
 which behaves like a monopole in the subspace
\foot{Note
that the difference between this monopole and the monopole solution given in
ref.\HaLi, where $\pa_4 A_\mu=0$ is required,
is that for the latter $r^2$ is not the same as that of the fivebrane,
but for the former it is so.}.

In the hyperbolic monopole case we required the rotational symmetry in the
$uv$-plane. We can also
solve the SDYM equation with $S^2$ rotational symmetry.
Let
\eqn\escoi{x^2=\rho\sin\theta\cos\varphi,\ x^3=\rho\sin\theta\sin\varphi,
\ x^4=\rho\cos\theta.}
In this coordinate system now the YM vector fields are
\eqn\yrmv{A_\mu dx^\mu=A_1 dx^1+A_\rho d\rho
+ \rho A_\theta d\theta + \rho\sin\theta A_\varphi d\varphi,}
where we can introduce a scalar field
$\Phi\equiv\rho A_\theta$.\foot{ Note that the
identification of the scalar fields are different in the $S^1$- and $S^2$-
symmetric cases.}
Then the SDYM equations, requiring that the fields do not depend on the angular
variables, reduce to
\eqn\resdym{F_{x\rho}=0,\ \ D_x\Phi=0=D_\rho\Phi,\ \ \ A_\varphi=0.}
This is rather a trivial system, but it still provides a source for a
fivebrane. The solutions of these equations are pure gauge solutions, which can
be gauged away so that we can set all the gauge fields zero. Then the anomaly
equation eq.\anoe\ simply becomes $\d H=0$. Thus we recover the elementary
fivebrane solution\DuLu\CaHaSt.

\newsec{(Euclidean) Sine-Gordon Case}

\def\zb{{\overline z}}
\def\wb{{\overline w}}
The (anti)self-dual YM equations have an interesting reduction to the
two-dimensional solitonic system, namely the sine-Gordon equation.
Here we shall attempt a new reduction of the (A)SDYM equation to the
Euclidean sine-Gordon equation for the gauge group SU(2) and the
Euclidean signature, then to solve the anomaly
equation eq.\anoe\ for this solution. The usual sine-Gordon system can be
recovered by further reducing this system,
incorporating the time dimension in $M^{1,5}$.

For the Euclidean signature we can introduce two sets of complex coordinates
for convenience, although one can use the real coordinates, as
\eqn\comcor{z=x+iy,\ \zb=x-iy,\ \ w=u+iv,\ \wb=u-iv,}
where $(x,y,u,v)$ are the cartesian coordinates.
In this coordinate system the SDYM equations will be written as
\eqn\sdymea{F_{z\zb}-F_{w\wb}=0,\ \ F_{z\wb}=0,\ \ F_{\zb w}=0,}
while the ASDYM equations are
\eqn\asdymea{F_{z\zb}+F_{w\wb}=0,\ \ F_{zw}=0,\ \ F_{\zb \wb}=0.}

For the gauge group SU(2) with the  generators
$J_\pm={\textstyle{1\over {\sqrt 2}}}(J_1\pm iJ_2),\ \ J_3$, which are in the
adjoint representation such that $(J_a)_{bc}=-i\epsilon_{abc}$,
we can introduce an ansatz for the gauge fields as
\eqn\idgau{A_z=f_1 J_3,\ A_\zb=f_2 J_3,\ A_w=g_1 J_+ +g_3 J_-,
\ A_\wb=g_2 J_- + g_4 J_+.}
With such identifications the SDYM equations reduce to
\eqn\nwrds{\eqalign{f_1&=\pa_z\ln g_2=-\pa_z\ln g_4,\cr
			f_2&=-\pa_\zb\ln g_1=\pa_\zb\ln g_3,\cr
		0&=\pa_z f_2-\pa_\zb f_1 -g_1 g_2 +g_3 g_4,\cr}}
and the conditions that $\pa_\wb f_1=\pa_w f_2=0,
\pa_w g_2=\pa_\wb g_3, \ \pa_w g_4=\pa_\wb g_1$. The last conditions can be
simply satisfied by requiring two Killing symmetries along $(u,v)$ directions
such that none of the fields depend on the
$(u,v)$-coordinates. For the ASDYM equation we obtain more or less the same set
of equations.

Now defining
\eqn\ancon{g_1=-g_2=e^{-{i\over 2}\psi},\ g_3=-g_4=e^{{i\over 2}\psi},}
we obtain the Euclidean version of the sine-Gordon equation,
\eqn\esigo{\pa_z\pa_\zb\psi-2\sin\psi={\textstyle{1\over 4}}(\pa_x^2 +
\pa_y^2)\psi - 2\sin \psi=0.}
The above is related, redefining  $y=it$, to the $(m^2=8)$
sine-Gordon equation
\eqn\msge{(\pa_t^2-\pa_x^2)\varphi+{m^2\over\lambda}\sin
\lambda\varphi=0,}
where  the coupling constant  $\lambda$ can be
rescaled away since we are not interested in quantizing this system  here.

In this background the anomaly equation eq.\anoe\ becomes up to the first order
of $\alpha'$
\eqn\sganoe{(\pa_x^2+\pa_y^2) \e^{-2\phi}=4\alpha'\left[\sin\psi\left(
\pa_x^2+\pa_y^2\right)\psi+\cos\psi\left((\pa_x\psi)^2+(\pa_y\psi)^2\right)
\right].}
Using the above sine-Gordon equation we can easily solve this equation to
obtain a solution
\eqn\bisgol{\e^{-2\phi}=\e^{-2\phi_0}+4\alpha'(1-\cos\psi),}
where $\psi$ satisfies the sine-Gordon equation
and $\phi_0$ is the value of the dilaton $\phi$ at
$x, y=\pm\infty$.

Due to the Derrick's theorem\ref\Derr{G.H. Derrick, J. Math. Phys. {\bf 5}
(1964) 1252\semi R. Hobart, Proc. Phys. Soc. {\bf 82} (1963) 201.}\
applied to the Euclidean sine-Gordon theory, there
is no finite-action static solution for $\psi$. Nevertheless, we can have
infinite-action static solutions, which do not generate any tunnelling effect.
In fact we can easily find the following solution:
\eqn\sgsol{\psi=4Q\tan^{-1}\left[\gamma\e^{\alpha x+\beta y}\right],}
where $\gamma$ is an arbitrary irrelevant constant so that we can set
$\gamma=1$
without loss of generality, and ${\alpha^2+\beta^2}=8$.
$Q=\pm 1$ is the soliton charge.
This solution is related to the soliton solutions of
the sine-Gordon equation eq.\msge,
\eqn\sgkink{\varphi=4Q\tan^{-1}\left[\exp\,
m{x-ct\over\sqrt{1-c^2}}\right],}
identifying
\eqn\resg{y=it,\ \ c=i\tilde{c},\ \ \alpha={m\over\sqrt{1+{\tilde c}^2}},\ \
\beta={m\tilde{c}\over\sqrt{1+{\tilde c}^2}},\ \ m=2\sqrt 2.}

It is straightforward to
show that the corresponding action of the Euclidean sine-Gordon
theory is indeed infinite for these solutions. However, this cannot be a
reason to abandon these solutions for our purpose because this action is not an
essential ingredient for fivebrane solutions.
 Note that the SDYM equation is not an
equation of motion so that the action for any reduced system from the SDYM
equation is not relevant to us.
Due to the self-dual YM
structure, the corresponding fivebrane solutions can still
saturate the necessary Bogomol'nyi bound for the energy density.
Strictly speaking, the fivebrane is not an instanton related to the tunnelling
effect because we work on the $(1+9)$ dimensional spacetime.
{}From this point of view, whether the action of the heterotic string is finite
or not is not really a relevant issue to us. We are just interested in looking
for some solitonic solutions.

Using eq.\sgsol,  now the dilaton eq.\bisgol\ becomes
\eqn\bslo{\e^{-2\phi}=\e^{-2\phi_0} + 16\alpha' {\e^{2(\alpha x+\beta y)}\over
\left(\e^{2(\alpha x+\beta y)}+1\right)^2}.}
Note that this solution does not have any singularity and depends on the
$x,y$-coordinates explicitly, not just on $x^2+y^2$.
This dilaton solution does not care about
the sign of the soliton charge $Q=\pm 1$, while the YM fields depend on the
charge $Q=\pm 1$.
We can also express the YM fields eq.\idgau\ in terms of eq.\sgsol\ as follows:
\eqn\ymsol{\eqalign{A_z&=-Q(\beta+i\alpha)
{\e^{(\alpha x+\beta y)}\over \e^{2(\alpha x+\beta y)}+1 } J_3,\cr
A_\zb&=Q(\beta-i\alpha)
{\e^{(\alpha x+\beta y)}\over \e^{2(\alpha x+\beta y)}+1 } J_3,\cr
A_w&={1-\e^{2(\alpha x+\beta y)}-i2Q\e^{(\alpha x+\beta y)}\over
	\e^{2(\alpha x+\beta y)}+1}J_+ +
{1-\e^{2(\alpha x+\beta y)}+i2Q\e^{(\alpha x+\beta y)}\over
	\e^{2(\alpha x+\beta y)}+1}J_- ,\cr
A_\wb&={1-\e^{2(\alpha x+\beta y)}-i2Q\e^{(\alpha x+\beta y)}\over
	\e^{2(\alpha x+\beta y)}+1}J_- +
{1-\e^{2(\alpha x+\beta y)}+i2Q\e^{(\alpha x+\beta y)}\over
	\e^{2(\alpha x+\beta y)}+1}J_+ .\cr}}
The fact that there are all the four dimensional YM fields indicates that the
solutions we have here are still  fivebrane solutions.

Now let us count the zero modes.
In the two-dimension parametrized by $(x, y)$ coordinates the soliton solutions
eq.\sgsol\ generate four zero modes, which are two for the two translational
symmetries of the $x, y$-directions, one for the $(\alpha^2+\beta^2=8)$
``scaling" symmetry and one for the O(2) rotational symmetry of $(\alpha
x+\beta
y)$. This last O(2) symmetry is due to the fact that the O(2) rotation of
$(x,y)$ can be compensated by O(2) rotation of $(\alpha, \beta)$.
Since the two Killing symmetries, $(\pa_u,\ \pa_v)$,
generate four extra zero modes for the
fivebrane, the fivebrane solution still has 120 bosonic zero modes, including
112 zero modes due to $E_8\to SU(2)\times E_7$, like in the ``gauge" solution
case. We expect that the fermionic zero modes counting is also similar to the
``gauge" solution case.

Note that the time-independent part of the sine-Gordon system can be easily
obtained by further imposing one more Killing symmetry, incorporating the
time-dimension from $M^{1,5}$. The corresponding fivebrane solutions can be
easily reduced from the Euclidean case.

\newsec{$\Phi^4$ case}

After we obtained the sine-Gordon system, we can easily reduce the previous
system to the $\Phi^4$ system, which has different type of
kink solutions in (1+1)-dimensional case.
In the Euclidean case we can again obtain analogues of these solutions,
though they are again infinite-action solutions.

For this purpose we use
\eqn\phfou{g_1=g_2=\e^{-i{\sqrt\lambda\over 2}\Phi},\ \
           g_3=g_4=\e^{i{\sqrt\lambda\over 2}\Phi},}
for the same ansatz eq.\idgau\
and by truncating at the leading order of $\lambda$, we obtain the field
equation for the $\Phi^4$ scalar field theory as
\eqn\phfoe{(\pa_x^2+\pa_y^2)\Phi+8\Phi-{\textstyle{4\over 3}}
\la\Phi^3=0.}

Again solving now the familiar anomaly equation, we obtain the dilaton
\eqn\phfdil{\e^{-2\phi}=\e^{-2\phi_0}+2\alpha'\lambda\Phi^2\left(1-{\lambda\over
12}\Phi^2\right),}
and the YM fields
\eqn\hhjjym{\eqalign{A_x&=-{\sqrt\la\over 2}\pa_y\Phi J_3,\ \ \ \ \ \ \ \ \
\ \ \ \ \		A_x={\sqrt\la\over 2}\pa_x\Phi J_3,\cr
	A_u&=\sqrt{2\la}\left(\Phi-{\la\over 24}\Phi^3\right) J_2,\
	A_v=i2\sqrt{2}\left(1-{\la\over 8}\Phi^2\right) J_1,\cr}}
where
\eqn\phkin{\Phi=Q\sqrt{{6\over\lambda}}{\rm tanh}(\alpha x+\beta y),\ \ Q=\pm
1,}
and $\alpha^2+\beta^2=4$.

Note that although eq.\phfdil\ does not depend on $\la$ explicitly for
solutions eq.\phkin, the results we have make sense only for small $\la$.

\newsec{Instanton Membrane}

The solutions we have derived so far are new fivebrane solutions.
On the contrary, the
instanton solutions of the self-dual YM equation in general lead to
$\pa_\theta A_\mu\neq 0$ for the coordinates given by eq.\ecor,
even though the relevant ingredients to construct the
fivebrane solution, namely, $\phi, \ H_{\mu\nu\la}$,
are independent from $\theta$. This raises a question speculated in ref.\Stro,
that is, keeping the instanton structure in the transverse space, whether we
can still reduce it to lower dimensional objects.

For the same three-dimensional subspace,
now if we do not require $\pa_\theta A_\mu=0$, we get
the structure suggested by Strominger, that is, the instanton lies in one
``internal'' and three ``external'' dimensions. But this case does not quite
define a compactified internal space.
Nevertheless, we can attempt to define resonable subspaces, which can be more
or less independent from the rest of the space. Among others\foot{Most of them
are rather physically awkward.}, there is at least one interesting case.
Note that $M^4$ is conformally equivalent
to $M^1\times S^3$. Using this,  we can get a membrane out of a
fivebrane in a $(1+3)$-dimensional subspace.
In this case the instanton lies in three ``internal'' dimensions.

Let us introduce the four dimensional radial coordinates as
\eqn\eccoi{\eqalign{x^1 &=r\sin\zeta\sin\theta\cos\varphi,\cr
x^2 &=r\sin\zeta\sin\theta\sin\varphi,\cr
 x^3 &=r\sin\zeta\cos\theta,\cr
 x^4 &=r\cos\zeta,\cr}}
and
$$r=a\e^{\xi/a},$$
then the metric for $M^1\times S^3$ is given by
\eqn\escmr{ds_4^2=\e^{-2\phi}\e^{2\xi/a}\left[d\xi^2+a^2
\left(d\zeta^2+\sin^2\zeta d\theta^2
+\sin^2\zeta\sin^2\theta d\varphi^2\right)\right],}
where $a$ is the radius of $S^3$ and the conformal factor depends only on
$\xi$.

We can construct a metric for (1+3)-dimensional subspace as
\eqn\fimrt{ds_{(1,3)}^2=-dt^2+(dx^5)^2+(dx^6)^2
+d\eta^2,}
where a new coordinate $\eta$ is introduced such as $d\eta^2=
\e^{-2\phi}\e^{2\xi/a} d\xi^2$, since this conformal factor
 is a function of $\xi$ only. In this case the ``internal'' space has a
so-called ``warp factor'' in contrast to the cases of known examples, where the
warp factor depending on the internal space appears in front of the space-time
metric\ref\warfac{P. van Nieuwenhuizen and N.P. Warner, Comm. Math. Phys. {\bf
99} (1985) 141\semi B. de Wit, D.J. Smit and N.D. Haridass, Nucl. Phys. {\bf
283} (1987) 165.}.

Though the vector potential $A_\mu$ depends on $\theta$-coordinate,
it is not surprising
because the instanton lies partly over $S^3$.
For small $a$ we can in principle
reduce the instanton solution in $M^4$ down to $M^1$
by doing a harmonic expansion over $S^3$.
In general there are nonvanishing massive modes.

\newsec{Discussion}

One of the
most important issues to make the fivebrane solution of heterotic string
theory really relevant to the physical world is to investigate what to happen
after the compactification from ten-dimension. Since the fivebrane is
dimensionally to large to fit into the four-dimension naturally, we should
expect that any compactified solution ever exists, it should show up as some
lower dimensional objects which could fit into the four-dimension.

In general one could expect that the fivebranes could appear as particles,
strings or membranes in the $(1+3)$ dimensional spacetime.
In this letter we have presented an explicit construction of solitonic
solutions
in the lower dimensional subspace of the transverse space $M^4$,
reduced from the heterotic fivebrane solutions in ten-dimension.
These solitons can survive after, for example, toroidal compactifications,
because they do not depend on the other coordinates explicitly.

As given in sect.6, the
example of putting instanton partially into the internal space
suggests that other compactified solutions of fivebranes would have
space-dependent warp factors in the metric of internal space. For example,one
can in principle map
$M^{1,9}$ conformally to $M^{1,3}\times S^2\times S^4$ or
$M^{1,3}\times S^3\times S^3$, we get a string or a membrane respectively.
Note that both internal spaces are K\"ahler and the sizes of the internal
spaces are controlled by two radii now.
Perhaps to confirm whether these really allow
compactified fivebrane solutions may not be an easy exercise.

We also expect that
the origin of the electric-magnetic duality in four-dimensional
world is originated from the string-fivebrane duality in ten-dimension in such
a way that the solitonic sector of the four-dimensional effective field theory
might be coming from the fivebrane sector of the string theory.

\bigbreak\bigskip\bigskip\centerline{{\bf Acknowledgements}}\nobreak

\par\vskip.3truein

The author would like to thank M. Duff, D. Nanopoulos and E. Sezgin
for conversations on related subjects.
This work was supported in part by NSF grant PHY89-07887 and a World Laboratory
Fellowship.


%
\listrefs
\vfill\eject
\bye